\documentclass[a4paper,12pt, epsfig]{article}
\usepackage{epsfig}
\usepackage{amssymb}
\usepackage{amsfonts}
\usepackage{amsmath}
\usepackage{graphicx}

\newskip\humongous \humongous=0pt plus 1000pt minus 1000pt

\newif\ifdtup

\catcode`\@=11

\@addtoreset{equation}{section}
\def\theequation{\thesection.\arabic{equation}}

\def\@normalsize{\@setsize\normalsize{15pt}\xiipt\@xiipt
\abovedisplayskip 14pt plus3pt minus3pt%
\belowdisplayskip \abovedisplayskip
\abovedisplayshortskip \z@ plus3pt%
\belowdisplayshortskip 7pt plus3.5pt minus0pt}

\def\small{\@setsize\small{13.6pt}\xipt\@xipt
\abovedisplayskip 13pt plus3pt minus3pt%
\belowdisplayskip \abovedisplayskip
\abovedisplayshortskip \z@ plus3pt%
\belowdisplayshortskip 7pt plus3.5pt minus0pt
\def\@listi{\parsep 4.5pt plus 2pt minus 1pt
     \itemsep \parsep
     \topsep 9pt plus 3pt minus 3pt}}

\relax

\catcode`@=12

\catcode`\@=11

\def\section{\@startsection{section}{1}{\z@}{3.5ex plus 1ex minus
   .2ex}{2.3ex plus .2ex}{\large\bf}}

\def\thesection{\arabic{section}}
\def\thesubsection{\arabic{section}.\arabic{subsection}}

\def\appendix{\setcounter{section}{0}
 \def\thesection{Appendix \Alph{section}}
 \def\thesubsection{\Alph{section}.\arabic{subsection}}
 \def\theequation{\Alph{section}.\arabic{equation}}}

\def\SymBoxes#1#2#3#4{\newdimen\un@t \un@t#3%
\raisebox{#1}{\rule{#2\un@t}{#4}\hskip-#2\un@t
\@tempdimb\un@t \advance\@tempdimb by-#4\@tempcntb#2\relax%
\@whilenum{\@tempcntb>0}\do{
\rule{#4}{\un@t}\hskip\@tempdimb \advance\@tempcntb by\m@ne}%
\hskip-#2\un@t \rule[\un@t]{#2\un@t}{#4}%
\rule[\un@t]{#4}{#4}\hskip-#4
\rule{#4}{\un@t}}\hskip-#4}                

\begin{document}

\newcommand{\beq}{\begin{equation}}
\newcommand{\eeq}{\end{equation}}
\newcommand{\bea}{\begin{eqnarray}}
\newcommand{\eea}{\end{eqnarray}}
\newcommand{\beas}{\begin{eqnarray*}}
\newcommand{\eeas}{\end{eqnarray*}}
\newcommand{\defi}{\stackrel{\rm def}{=}}
\newcommand{\non}{\nonumber}
\newcommand{\bquo}{\begin{quote}}
\newcommand{\enqu}{\end{quote}}
\renewcommand{\(}{\begin{equation}}
\renewcommand{\)}{\end{equation}}
\def \eqn#1#2{\begin{equation}#2\label{#1}\end{equation}}
\def\IZ{{\mathbb Z}}
\def\IR{{\mathbb R}}
\def\IC{{\mathbb C}}
\def\IQ{{\mathbb Q}}
\def\de{\partial}
\def\Tr{ \hbox{\rm Tr}}
\def\H{ \hbox{\rm H}}
\def\HE{ \hbox{$\rm H^{even}$}}
\def\HO{ \hbox{$\rm H^{odd}$}}
\def\K{ \hbox{\rm K}}
\def\Im{ \hbox{\rm Im}}
\def\Ker{ \hbox{\rm Ker}}
\def\const{\hbox {\rm const.}}
\def\o{\over}
\def\im{\hbox{\rm Im}}
\def\re{\hbox{\rm Re}}
\def\bra{\langle}\def\ket{\rangle}
\def\Arg{\hbox {\rm Arg}}
\def\Re{\hbox {\rm Re}}
\def\Im{\hbox {\rm Im}}
\def\exo{\hbox {\rm exp}}
\def\diag{\hbox{\rm diag}}
\def\longvert{{\rule[-2mm]{0.1mm}{7mm}}\,}
\def\a{\alpha}
\def\dag{{}^{\dagger}}
\def\tq{{\widetilde q}}
\def\p{{}^{\prime}}
\def\W{W}
\def\N{{\cal N}}
\def\hsp{,\hspace{.7cm}}
\newcommand{\C}{\ensuremath{\mathbb C}}
\newcommand{\Z}{\ensuremath{\mathbb Z}}
\newcommand{\R}{\ensuremath{\mathbb R}}
\newcommand{\rp}{\ensuremath{\mathbb {RP}}}
\newcommand{\cp}{\ensuremath{\mathbb {CP}}}
\newcommand{\vac}{\ensuremath{|0\rangle}}
\newcommand{\vact}{\ensuremath{|00\rangle}                    }
\newcommand{\oc}{\ensuremath{\overline{c}}}
\begin{titlepage}
\begin{flushright}
\end{flushright}
\def\thefootnote{\fnsymbol{footnote}}

\begin{center}
{\Large {\bf
Tomograms of Spinning Black Holes\\
\vspace{0.3cm}
}}
\end{center}

\bigskip
\begin{center}
{\large  Chethan
KRISHNAN}\\
\end{center}

\renewcommand{\thefootnote}{\arabic{footnote}}

\begin{center}
{\em  { SISSA,\\
Via Beirut 2-4, I-34014, Trieste, Italy\\
{\rm {\texttt{krishnan@sissa.it}}}\\}}

\end{center}

\noindent
\begin{center} {\bf Abstract} \end{center}
The classical internal structure of spinning black holes is vastly different from that of static black holes. We consider spinning BTZ black holes, and probe their interior from the gauge theory. Utilizing the simplicity of the geometry and reverse engineering from the geodesics, we propose a thermal correlator construction which can be interpreted as arising from two entangled CFTs. 
By analytic continuation of these correlators, we can probe the Cauchy horizon. Correlators that capture the Cauchy horizon in our work have a structure closely related to those that capture the singularity in a non-rotating BTZ. As expected, the regions beyond the Cauchy horizon are not probed in this picture, protecting cosmic censorship.

\begin{center}
Quantum Field Theory in Curved Spacetime.
\end{center}

\vfill

\end{titlepage}
\bigskip

\hfill{}
\bigskip

\tableofcontents

\setcounter{footnote}{0}
\section{\bf Introduction}

\noindent
In general relativity, Cauchy horizons are the boundaries of regions of
spacetime that lie outside the Cauchy development of an initial value surface\footnote{See Appendix A for a self-contained review of the relevant ideas.}. In other words, the initial value data on a
Cauchy surface is not enough to determine what happens beyond its future
Cauchy horizon. It turns out that the inner horizons of
charged/rotating black holes are Cauchy, and one reason why they are
perplexing is because they are surfaces of infinite blue-shift: a mode
that is regular at the outer horizon, will undergo a mass inflation when
it
reaches the inner horizon, and will have a divergent stress-energy tensor
\cite{inner, Bala}.
This means that there is no possibility of treating any mode as a ``small"
perturbation, because the backreaction on the geometry can never be
ignored. Ref. \cite{Brady:1995un} contains numerical simulations and
analytical
studies on the evolution of the perturbed spacetime containing a Cauchy
horizon.

The existence of the Cauchy horizon also gives rise to some tension with
the Cosmic
Censorship Conjecture. If we take the definition of
censorship to
imply that any inextendible spacetime that contains only physically
reasonable matter {\em must} be globally hyperbolic,
then it turns out that
a blackhole spacetime with an inner horizon will violate the conjecture.
But if we impose the condition that the spacetime also be {\em generic} \cite{Clarke:1994cw}, then that will save Censorship: the inner
horizon is not generic, because it is unstable against perturbations. More
about cosmic censorship
and black-hole horizons in the context of cosmological spacetimes can be
found in \cite{cosmo}.

Since the AdS/CFT correspondence provides us with a dictionary connecting
geometry and gauge theory, one might wonder whether it is possible to
understand the Cauchy horizon in the dual gauge theory. Indeed, right
after the original AdS/CFT proposal, the thermal properties of black
holes in $AdS_5\times S^5$ were related to thermal effects in
the dual ${\mathcal N}=4$
supersymmetric Yang-Mills theory \cite{Witten:1998zw}. But it was not fully clear
how much of the internal structure of the black hole is captured  in the gauge theory.
A proposal was made by Maldacena in \cite{Maldacena:2001kr, Israel:1976ur} for studying the fully extended geometry of large AdS black holes by identifying the Hartle-Hawking vacuum
with an entangled state in the CFT.
Using this idea, various efforts have been made to see beyond the horizons of static black holes using boundary correlators \cite{
Hemming:2002kd, Shenker1, Shenker2, Levi,
Bala}. In particular, it was found by
\cite{Shenker1} that the
analyticity properties of the coordinate space correlators encode the
physics beyond the horizon. This strategy has been applied also for spacetimes with Cauchy horizons: for rotating BTZ black holes, it was used in \cite{Levi, Bala}. 

A very useful approach was developed by Shenker and collaborators in \cite{Shenker1, Shenker2} to probe
beyond the horizon using spacelike
geodesics connecting the two asymptotic boundaries of AdS black holes. In the
large mass limit, these geodesics
are the primary contributions to the two-point correlation
functions in the boundary quantum field theory. (See \cite{Brecher:2004gn} for
the investigation of such geodesics in the context of charged AdS black holes.). Festuccia and Liu \cite{FL1} put this construction on a more concrete footing by explicitly developing the map between correlators in the Hartle-Hawking (HH) vacuum, and geodesics and saddle points. They did this for the AdS Schwarzschild black hole. Using the analyticity properties of momentum-space
correlators, they identified the signatures of the singularity in the gauge theory and observed the holographic generation of
``time" inside the event horizon. In this paper, adopting the viewpoint that geodesics are paramount in defining the entangled CFT correlators, we use these methods to study the rotating BTZ black hole and its Cauchy horizon. The construction of HH-like thermal states for spinning black holes is a subject that is not fully under control \cite{Frolov, Kay}, but we will see that for spinning BTZ, there does  exist a vacuum (more precisely, a construction of two-point functions for our purposes) that is reasonable from the geodesic perspective. The Penrose diagram of spinning black holes involves an infinite number of asymptotic regions, but we will present some arguments \cite{Maldacena:2001kr} that to define the theory from the boundary, it is most reasonable to work only with two of them.

One advantage of the BTZ black hole over the higher dimensional Reissner-Nordstrom-AdS black hole or Kerr-AdS black hole is that the scalar field
equation admits an exact solution, so we do {\em not} need an analysis of
the quasinormal modes in order to (approximately) determine the poles of
the Green functions. This is very handy because the  poles and the
analyticity structure of the Green functions will be crucial for our analysis. But along with this simplification comes the drawback that the BTZ singularity is an orbifold, so some restraint needs to be exhibited in drawing messages for higher dimensions from our results. We discuss these things in more detail in the final section.

One claim that arises from our investigation is that the Cauchy
horizon of the rotating BTZ black hole has a
signature in the thermal correlators that is the same as
that of
the singularity of a static BTZ black hole (except for a simple variable redefinition). This could be an
indication
that the perturbed inner horizon eventually settles down and forms a
singularity. Note that because of the instability, spacetimes with Cauchy horizons are a ``measure-zero" subset of the space of solutions, and so it is natural that a quantum description will be dominated by the end-point of the instability. This is also further evidence that the inner horizon is
an artifact of the classical solution, and is suppressed in the full quantum theory. Conversely, if one takes at face value the tentative evidence from numerical relativity \cite{Brady:1995un} that the eventual fate of
the perturbed Cauchy horizon is a singularity, it is satisfying that the gauge theory (which, in light of the AdS/CFT correspondence is in principle the {\em definition} of quantum gravity with AdS boundary conditions) is consistent with such an interpretation. As a spinoff, our results agree also with the observation in the literature \cite{Levi, Bala} that the regions beyond the Cauchy horizon are not visible in the gauge theory\footnote{There is a technical caveat to this statement, which we will clarify later on.}. Thus the cosmic censor is still at work, and we are saved from timelike singularities and new asymptotic regions. The final result is self-consistent with our starting fact that we defined our theory in terms of only two asymptotic regions, when we considered the (entangled) CFTs that live on each boundary.

Some other perspectives on our results are provided in the concluding section.

The organization of the paper is as follows. In the next section, we work out the geodesics in the spinning BTZ spacetime, show it has an intuitive variable redefinition, and demonstrate its geometric origins. In section 3, we discuss the construction of thermal correlators for black hole spacetimes, and specifically for our spinning BTZ. In the process, we clarify some aspects of quantizing scalar fields in the BTZ geometry, including the choice of ensemble and the question of super-radiance. The heuristic construction of section 3 will be further justified in section 4, when we compute correlators and show that they are dominated by the geodesics of section 2, using a large mass WKB approximation developed by Festuccia and Liu. Along the way, we also make use of the exact solution of the wave equation with the appropriate boundary conditions. In the next section using all the ingredients from the previous sections, we discuss the correlators that probe the Cauchy horizon. The concluding section includes comments and open questions, including a discussion of generalizations to higher dimensions. Technical asides and some review-items are relegated to various appendices.

Related recent work having to do with black holes in AdS/CFT and string theory can be found in \cite{etc, Yang:2006pc}.

\section{Geometry and Geodesics}

The BTZ black hole is a black hole in 2+1 dimensions \cite{BTZ, BHTZ} and it is interesting because it exhibits many properties of 3+1 black holes, while still allowing us to do many exact computations. The simplicity arises because the geometry is simply an orbifold of $AdS_3$ and (as a corollary) the singularity is a delta function orbifold singularity and not a genuine curvature singularity. But like the 3+1 Kerr black hole, BTZ has horizons, it has an interesting thermodynamics and it can form as the endpoint of gravitational collapse \cite{Carlip2}. This last fact is interesting because it is an indication that we do not loose all of the interesting physics even though the singularity is an orbifold. The metric of the BTZ black hole in various coordinates, together with its conformal structure and some comments about the Euclidean section are provided in Appendix B. 

One of the results of \cite{Shenker1, Shenker2, FL1} is that (at least for static black holes) the dominant contribution to the boundary-boundary correlators comes from certain spacelike geodesics that go inside the bulk. The basic idea is that in the large mass limit of the scalar field (whose correlators we are computing), the propagation can be approximated as the classical motion of a massive particle along a geodesic. So we will start by studying such geodesics, as written in the standard BTZ coordinates.

\subsection{Spacelike Geodesics}

The spacetime has two Killing vectors,
$\zeta=\partial_t$ and $\chi=\partial_\phi$ in terms of the BTZ coordinates written down in Appendix B. We will call the conserved
quantities
associated with them energy $E$ and angular momentum $q$ respectively.
The
Killing vectors provide a recipe for writing down the first integrals of
motion. In our case, using
$\zeta^{\mu}=(1,0,0)$, and $\chi^\mu=(0,0,1)$, we can write down the
conserved quantities associated with them to be
\begin{eqnarray}
E\equiv-g_{\mu\nu}\zeta^\mu u^\nu
=(r^2-r_+^2-r_-^2)\frac{{\rm d}t}{{\rm d}\lambda}+r_-r_+\frac{{\rm
d}\phi}{{\rm d}\lambda}, \\
q\equiv g_{\mu\nu}\chi^{\mu} u^{\nu}=-r_-r_+\frac{{\rm d}t}{{\rm
d}\lambda}+r^2\frac{{\rm
d}\phi}{{\rm
d}\lambda}, \hspace{1.0cm}
\end{eqnarray}
where $u^\mu=\frac{{\rm d}x^\mu}{{\rm d}\lambda}$ and $\lambda$ is the
affine
parameter. Inverting these and solving for $\dot \phi$ and $\dot t$
gives,
\begin{eqnarray}
\label{t}\frac{{\rm d}t}{{\rm
d}\lambda}=\frac{Er^2-qr_-r_+}{(r^2-r_-^2)(r^2-r_+^2)}, \hspace{0.25 in} \label{num1}\\
\label{phi}\frac{{\rm d}\phi}{{\rm d}
\lambda}=\frac{q(r^2-r_-^2-r_+^2)+Er_-r_+}{(r^2-r_-^2)(r^2-r_+^2)}.\label{num2}
\end{eqnarray}
To get information from beyond the event horizon what we need are
spacelike geodesics because we will be taking correlators
between two different asymptotic boundaries \cite{Israel:1976ur,
Maldacena:2001kr}.
So the (integrated) geodesic equation for our choice
of signature is $g_{\mu\nu}u^\mu u^\nu=1$. Using the above expressions for
$\dot \phi$ and $\dot t$, we arrive at the explicit form,
\eqn{geodesic}{\Big(\frac{{\rm d}r}{{\rm d}
\lambda}\Big)^2+\frac{2Eqr_-r_+-q^2\big(r_+^2+
r_-^2\big)}{r^2}-\frac{(r^2-r_+^2)(r^2-r_-^2)}{r^2}+q^2=E^2.}
First thing we note about this equation is that unlike in the
case of the static black hole, there is an unavoidable mixing of $q$ and $E$, so it is not immediately possible to interpret $E$ as the energy of a particle in a potential. But one crucial observation is that if one works with two  new  linear combinations of the basic variables, 
\bea
E'=\frac{Er_+-q r_-}{\sqrt{r_+^2-r_-^2}} \label{redef1}\\
q'=\frac{Er_--q r_+}{\sqrt{r_+^2-r_-^2}} \label{redef2}
\eea
the equation can be brought to the form of an energy conservation equation for a particle in a well,
\bea
\frac{r^2}{(r^2-r_-^2)}\Big(\frac{{\rm d}r}{{\rm d}
\lambda}\Big)^2+(r^2-r_+^2)\Big(\frac{q^2}{r^2-r_-^2}-1\big)=E^2. \label{geodesic2}
\eea 
We have suppressed the primes in the new variables. The form of the potential is precisely what one would get for the static BTZ black hole, if one were to set $r_-=0$. This is an indication of many of the things that follow: we will often find that the location $r=r_-$ takes the place of the singularity, in the case of the spinning black hole. 
Clearly, after a minor redefinition of $r$, the equation above can be thought of as particle of energy $E^2$ 
in a potential given by,
\eqn{potential}{-V(r)=(r^2-r_+^2)\Big(1-\frac{q^2}{r^2-r_-^2}\big)
}
The plot of the potential (in fact $-V$) for some choices of parameters is in
fig. 2. When we make the connection between geodesics and correlators, the parameters $E, q$ will often be analytically continued. 
\begin{figure}
\begin{center}
\includegraphics[
]{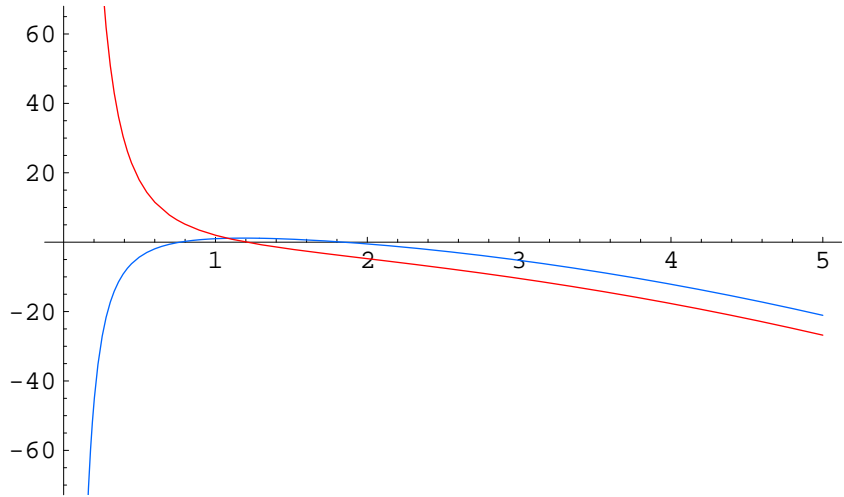}
\caption{
Schematic plots of (\ref{potential}) for real and imaginary $q$. 
The scales in the figure should not be taken seriously. }
\label{potplot}
\end{center}
\end{figure}

\subsection{New-BTZ Coordinates}

The variable redefinition we did in (\ref{redef1}-\ref{redef2}) is at the moment merely a matter of convenience. But it becomes much more suggestive when we realize that it follows immediately from a coordinate change of the BTZ metric:
\bea
t'=\frac{t-\Omega \phi}{\sqrt{1-\Omega^2}}, \ 
\phi '=\frac{\phi-\Omega t}{\sqrt{1-\Omega^2}} \label{new-transformation}
\eea
where $\Omega\equiv r_-/r_+$.
Under this coordinate transformation, the BTZ coordinates of Appendix B become
\bea
ds^2=-(r^2-r_+^2)dt^2+\frac{r^2 dr^2}{(r^2-r_+^2)(r^2-r_-^2)}+(r^2-r_-^2)d\phi^2, \label{newBTZ}
\eea
where again we have suppressed primes. We will call these coordinates the new-BTZ coordinates, and we will use them almost exclusively in what follows\footnote{We use the same notation $(t, r, \phi)$ for both BTZ and new-BTZ coordinates to avoid running out of symbols, but we will be careful enough to emphasize it when we use the former.}. 

A couple of comments are in order. Note that we are going to a frame co-rotating with the outer horizon as far as the $\phi$ coordinate is considered.  If we did only the $\phi$-shift, the new $\phi$ will still be periodic at fixed $t$ and this is a standard form for the metric, as is written (for example) in \cite{Bala}. But together, the $t$ and $\phi$ coordinate changes force a form of the BTZ quotienting (compare with (\ref{quotient})) that looks like 
\bea
(t, \phi) \sim \Big(t- \frac{2\pi\Omega}{\sqrt{1-\Omega^2}}, \phi+ \frac{2\pi}{\sqrt{1-\Omega^2}}\Big). \label{newquotient}
\eea
It can be checked that there are still no causal pathologies in the $t$ variable: if we were to draw the variables on a cylinder the identification looks roughly like a helix\footnote{I thank J. Evslin for a discussion on this.}. But the interpretation that a period of $\phi$ represents the horizon (at $r=r_+$) is no longer true.  In later sections, we will quantize scalar fields in the BTZ background and we will be concerned with high frequency massive modes which can be associated with classical particle propagation (basically (\ref{scaled}) with $\nu \rightarrow \infty$). This means that the modes are effectively continuous even though there is quantization arising from the periodicity. Note that these coordinates enable us to maximally take advantage of the fact that BTZ is locally $AdS_3$. 

Another useful observation is that as $r \rightarrow \infty$, the metric takes the usual Poincare patch form of $AdS_3$. 

We will have more to say about various aspects of this metric in later sections, but for now, we turn to the determination of Killing quantities and geodesics, as we did in the previous subsection. The relevant equations are immediately written down as
\bea
(r^2-r_+^2)\frac{d t}{d\lambda}=E, \ \ (r^2-r_-^2)\frac{d\phi}{d\lambda}= q, \hspace{0.35in} \label{first}\\ 
\frac{r^2}{(r^2-r_-^2)}\Big(\frac{{\rm d}r}{{\rm d}
\lambda}\Big)^2+(r^2-r_+^2)\Big(\frac{q^2}{r^2-r_-^2}-1\big)=E^2. \label{second}
\eea
The last equation was precisely what we got before for the radial geodesic equation, but now we see its origins in terms of the rewritten metric. The first two equations can also be easily seen to be identical to eqns (\ref{num1}, \ref{num2}).

Using the above relations, we can define the proper distance $L(E,q)$,
the proper time $t(E,q)$ and
the proper angular distance $\phi(E,q)$ between the initial and final
points
of a spacelike geodesic. We will write them down here explicitly for
future reference. To obtain $t(E,q)$ and $\phi(E,q)$, we solve for
${\rm d}\lambda$ from (\ref{geodesic}) and plug it into (\ref{t}),
(\ref{phi}):
\begin{eqnarray}
t(E,q)&=&2\int_{r_c}^{\infty}\frac{E \ r}{(r^2-r_+^2)\sqrt{(r^2-r_-^2)}}\frac{{\rm
d}r}{\sqrt{E^2+(r^2-r_+^2)-\frac{q^2(r^2-r_+^2)}{(r^2-r_-^2)}}}, \label{t-integ}\\
\phi(E,q)&=&2\int_{r_c}^{\infty}\frac{q \ r}{(r^2-r_-^2)^{3/2}}\frac{{\rm
d}r}{\sqrt{E^2+(r^2-r_+^2)-\frac{q^2(r^2-r_+^2)}{(r^2-r_-^2)}}} \label{phi-integ},
\end{eqnarray}
where $r_c$ is defined by $V(r_c)=E^2$, and the factor of 2 accounts for
the fact that the geodesic comes in from infinity, turns ``around" at the
turning point $r_c$ and then goes to the other asymptotic infinity. To
define $L(E,q)$ on the other hand, we choose the affine parameter as the
proper length and regularize by subtracting a $\log r$:
\eqn{L}{L(E,q)=2\lim_{r\rightarrow
\infty}\Big[\int_{r_c}^{r}\frac{r{\rm
d}r}{\sqrt{(r^2-r_-^2)}\sqrt{E^2+(r^2-r_+^2)-\frac{q^2(r^2-r_+^2)}{(r^2-r_-^2)}}}-\log r\Big].}
This regularization is explained (in the closely related context of higher dimensional AdS Schwarzschild black holes) in \cite{Shenker2}. We will later explicitly evaluate these integrals.

We will see that with the notion of geodesics defined here, it makes sense to talk about probing the Cauchy horizon from an entangled CFT, in the large-$\nu$ limit that we mentioned. We turn to a heuristic construction that can provide such a dual description next.

\section{Entangled CFTs and Thermal States}

In this section we will give a prescription for thermal correlators that can naturally give rise to the spacelike geodesics of the previous section.  
This will also have an entangled CFT interpretation according to Maldacena \cite{Maldacena:2001kr}. We start with some comments about vacuum states for scalar fields on bifurcate Killing horizons (in particular our geometry), the reader should consult \cite{Birrell, RossReview, Wald} for details on QFT on black hole backgrounds.

\subsection{A Thermal Correlator}

We take the geodesics (\ref{first})-(\ref{second}) as the fundamental objects. It can be seen that these geodesics can probe the inner horizon (we will see this in more detail in section 5). If the spinning black hole has an entangled CFT description, we expect that these geodesics have an interpretation in terms of thermal correlators.  Much of what we do in this section is based on heuristic generalizations of the Rindler wedge and on the idea that there should be an entangled description between two halves of the Rindler wedge. In particular, the arguments are meant to motivate, rather than derive our proposal.

There exists a general strategy for constructing thermal (Hartle-Hawking-like) vacua which is based on the fact that the spacetime under consideration has a bifurcate Killing horizon\footnote{All spacetimes we discuss have bifurcate Killing horizons. I thank A. Virmani for clarifications on some questions related to them.}. The basic idea is that such spacetimes can be mapped to the Rindler wedge where the problem has a well-known solution \cite{Birrell} in terms of Bogolubov transformations. In the large mass-large frequency limit which results in the geodesic approximation, we expect that the quantization of the modes due to the quotienting in (\ref{newBTZ}) can be ignored. (We work exclusively with momentum space correlators in this work.) 
When the quotienting is ignored, the metric (\ref{newBTZ}) can be thought of as that of a bifurcate Killing horizon, where $r=r_+$ is a bifurcation line (namely, $\phi$-coordinate). 
The standard strategy now to construct a thermal state is to write down the Kruskal-like coordinates for the metric in (\ref{newBTZ}). Then we will interpret the two Rindler halves that arise as the two asymptotic regions corresponding to the entangled CFTs. Using standard Bogolubov transformations, this would give us a construction of the putative vacuum state. We do this in an appendix. Essentially, this ends up giving us a canonical ensemble description of the thermal state\footnote{To be more precise, all we need is that our thermal state is one that gives rise to momentum space correlators that reduce to the canonical thermal correlators in the large $\nu$ limit.}, but note that the energy that one uses in this description is related by a variable redefinition (\ref{redef1}, \ref{redef2}) to the standard notion of energy and (angular) momentum in BTZ. 

We emphasize that this construction is a proposal and not a derivation. In particular, there is an order of limits ambiguity associated to ignoring the quotienting in the large-$\nu$ limit. But we will see that the results that one gets from this proposal are reasonable, so we will take this as our working hypothesis for defining thermal correlators for spinning BTZ in the large-$\nu$ limit. Note that effectively, our consruction involves entanglement between two Poincare $AdS_3$ patches in an appropriate way.

It is generally believed that spinning black holes should be described by grand canonical ensembles. But in the AdS-CFT context, the interpretation of the situation might be a bit different. We will see in subsection \ref{super} that to define a vacuum state for scalar fields which does not get into trouble with super-radiance, we need to quantize with respect to a time coordinate that is timelike everywhere outside the horizon. The time coordinate in a co-rotating (with the horizon) coordinate system is precisely such a coordinate. It seems natural from an AdS/CFT perspective that quantization in such a co-rotating frame (where the the chemical potential $\Omega$ is effectively invisible) should in fact be described in the canonical ensemble. We will make some comments in the final section about the precise interpretation of such a canonical thermal state for rotating black holes. Since our coordinate system is co-rotating, we believe that  a natural thermal state to consider is in the canonical ensemble. We will also see later on that when we choose the canonical form, it gives rise to very reasonable matches and interpretations. So it seems that especially in the specific case of BTZ, an appropriate canonical ensemble construction is useful.

The fundamental reason why this works computationally is because of the simple form of the BTZ metric. In fact by defining $\rho=\sqrt{r^2-r_-^2}$, we can bring the spinning BTZ metric (\ref{newBTZ}) to the form
\bea
ds^2=-(\rho^2-R^2)dt^2+\frac{d\rho^2}{\rho^2-R^2}+\rho^2 d\phi^2 \label{staticform}
\eea
where $R^2=r_+^2-r_-^2$.
This has the same form as that of the non-rotating BTZ metric, except for the fact that $\rho=0$ is classically the Cauchy horizon and {\em not} the singularity\footnote{Note also the quotienting (\ref{newquotient}).}. 
We are already seeing hints from the geometry that there exists a close analogy between the Cauchy horizon and the static BTZ singularity. 

The bottomline is that we take the construction presented in Appendix C 
as the definition (effective at least in the large $\nu$ limit) of our thermal correlators. This gives rise to the bulk correlator (see (\ref{separable}) and Appendix C for notation)
\bea
{\cal G}_+(r,r'; \omega, p)=\frac{1}{2\omega}\frac{e^{\beta \omega}}{e^{\beta \omega}-1}X_{\omega p}(r) X_{\omega p}(r')
\eea
Now, using standard Lorentzian\cite{LAdSCFT}-AdS/CFT\cite{AdSCFT}, we can define correlators of boundary operators that are dual scalar fields in the bulk. The real-time thermal correlators between these boundary operators can be defined by taking the bulk correlators in the limit when the insertions go to the boundary. We will deal with boundary Wightman correlators following the conventions of \cite{FL1}. These can be constructed as
\bea
G_+(\omega,p)=\lim_{r\rightarrow \infty} (2\nu r^\Delta)(2 \nu {r'}^{\Delta}){\cal G}_+(r,r'; \omega, p) \label{boundarylimit}
\eea
where in three dimensions, for a scalar field of mass $m$,  
\eqn{dimensions}{\Delta_{\pm}=1\pm\nu, \ \ {\rm with} \ \
\nu=\sqrt{1+m^2}.}

\subsection{Two CFTs for One Black Hole}

The reason we are interested in the thermal vacuum is because it shows up in a proposal by Maldacena \cite{Maldacena:2001kr} for describing AdS black holes as thermal states using two entangled CFTs. The correlators in the thermal vacuum will be interpreted as correlators in the entangled CFT.

The first step in the Maldacena construction is the observation (made earlier by Israel) that a thermal state can be written as an entangled state. A ``thermal state" $|0(\beta, \mu)\rangle$ in the grand canonical ensemble 
is defined as a state where the relation
\bea \label{thermal}
\langle 0(\beta,\Omega)| A |0(\beta,\Omega)\rangle = \frac{1}{Z}{\rm Tr}(e^{-\beta (H-\Omega J)} A)
\eea
holds for any observable $A$. The $\Omega$ acts as the chemical potential for the appropriate charge, $J$. Here $Z$ is the grand-canonical partition function
\bea
Z={\rm Tr}(e^{-\beta (H-\Omega J)}).
\eea
The fact that a state is thermal means that we don't have a complete description of it, and that our description coarse-grains (i.e., traces) over some degrees of freedom. This is the ``entangled'' description of a thermal state.
The key observation is that (\ref{thermal}) can be reproduced by explicit computation if we define
\bea
\label{BHstate}
| 0(\beta,\Omega)\rangle = \frac{1}{Z[\beta,\Omega]^{\frac{1}{2}}} \sum_{n} e^{-\frac{\beta\, (E_n-\Omega J_n)}{2} } |E_n, J_n\rangle
\otimes |E_n, J_n\rangle\,
\eea
if $A$ acts only on one of the Hilbert spaces in the tensor product.

The maximally extended spacetime of a black hole in AdS involves (at least) two boundaries. So it is natural to propose that the entangled state (\ref{BHstate}) between the CFTs living on these two boundaries is precisely the (Hartle-Hawking) thermal state of the black hole \cite{Maldacena:2001kr}. The standard AdS-CFT correspondence allows us to compute CFT Green functions as boundary limits of bulk correlators. So we can compute the Green's functions in the Hartle-Hawking vacuum, take the limit where the points go to the (same) boundary and get the thermal Green functions in the CFT. If we take the points to go to distinct boundaries then we get thermal correlators where one operator is inserted on one CFT and the other is inserted on the other CFT. These correlators correspond to geodesics that go inside the horizon, and they are what we will use to extract the information we are after.

When the angular momentum $J$ of a black hole becomes non-zero, its conformal structure undergoes a drastic change: in particular, the Penrose diagrams of rotating black holes can be extended infinitely. They involve new asymptotic boundaries, Cauchy horizons, and timelike singularities (instead of the spacelike singularities in the non-rotating case). So in the grand canonical version of the Maldacena prescription outlined above, most of these regions have to be (implicitly) considered fictitious\footnote{Note that when the chemical potential is zero, the black hole becomes uncharged and static, and one can work with the canonical ensemble. Because there are only two asymptotic boundaries, things are straightforward.}. Note that in our canonical ensemble formulation of the previous subsection, such problems are automatically avoided: we can put the two CFTs on the boundaries of the two Rindler-like half-spaces from the previous subsection. 
Putting the CFTs on only two boundaries is natural from the above entangled/thermal description, as well as the philosophy that the boundary theory (albeit thermal) is the full  definition of the theory. Conversely, we can take such a description as the definition of the theory and aim to see what this view implies for the internal structure of the black hole. At the end of our work, we will find evidence that this is in fact a self-consistent assumption: the regions beyond the Cauchy horizon will in fact be excised from the CFT correlators. This is also what one expects from numerical gravity simulations. 

Another reason why it is reasonable to consider only two boundaries seriously in the CFT picture is that much of the motivation for the Maldacena proposal comes from the Hartle-Hawking construction of the wavefunction of the eternal black hole Universe\footnote{Not to be confused with the Hartle-Hawking thermal state \cite{Hartle1}. The latter is a state defined on the Fock space of fields in the black hole geometry, while the former is a wave-function (in the Wheeler-DeWitt sense) of the spacetime itself. When the backreaction of the scalar on the geometry can be ignored, the only contribution to the full wave function that we will need to keep track of is the scalar part. This is the premise of QFT in curved space.}\cite{Hartle2}. There, two boundaries are the only possibility.
The conclusions of this paper show that the Cauchy horizon looks like a singularity from the boundary theory, so that the Maldacena proposal is indeed self-consistent.

\subsection{Super-radiance and Stability}\label{super}

Spinning black holes can exhibit super-radiance which can make the vacuum states defined on them ill-defined \cite{Kay}. In our construction, we could bypass this problem by going to a new coordinate frame because we are in AdS, but it is instructive to see this in detail. The basic idea behind super-radiance is easy to understand. When a mode of energy $\delta M\sim \hbar \omega$ and angular momentum $\delta J\sim \hbar k$ is incident on a black hole, the hole's mass and angular momentum change according to $dM/dJ= \omega/k$. This, together with the first law $dM=T_H dA +\Omega dJ$ implies that $dM=T_H \omega dA/(\omega-k \Omega)$. Since $dA \ge 0$ by the second law, this means that for incident modes that satisfy $\omega<  k\Omega$,
the hole looses mass. This is the wave-analogue of the Penrose process. The scattered wave has more energy than the incident wave because the hole is loosing its energy to the mode, and this is the phenomenon of super-radiance.

Super-radiance arises because rotating black holes have an ergosphere that stretches outside the outer horizon from where angular momentum can be extracted without falling into the hole.  This is because the energy as defined from infinity is with-respect to a time-like Killing vector that is no longer timelike inside the ergosphere. If there exists a globally defined time-like rotational Killing vector that stretches all the way to the boundary of the spacetime, there will not be any super-radiant instabilities. For Kerr black holes in AdS, such an argument was made in \cite{Reall} for black holes that are not rotating too fast. We will adapt that argument here for the case of BTZ. From the metric presented in (\ref{BTZmetric}), we can write down an angular velocity at the horizon given by\footnote{See Appendix B for notations. We are working with the usual BTZ coordinates here.}
\bea
\Omega=\frac{r_-}{r_+}.
\eea
Now the co-rotating Killing field is
\bea\label{chi}
\chi=\frac{\partial}{\partial t}+\Omega\frac{\partial}{\partial \phi}.
\eea
It can be checked directly using the metric, that the norm of this vector is given by
\bea
||\chi||^2=-\frac{(r^2-r_+)^2(r_+^2-r_-^2)}{r_+^2}.
\eea
So the vector is everywhere timelike outside the outer horizon. Using this $\chi$ we can repeat the argument presented in section II.B of \cite{Reall}. This shows that the BTZ spacetime is stable against superradiance, at least away from extremality.

To get a time coordinate which can be used to define positive energy modes, we can set $\chi=\frac{\partial}{\partial T}$.  Writing (\ref{chi}) as 
\bea
\frac{\partial}{\partial T}=\frac{\partial t}{\partial T}\frac{\partial}{\partial t}+\frac{\partial \phi}{\partial T}\frac{\partial}{\partial \phi},
\eea
which implies that the new coordinates $T$ and $\Phi$ have to be defined by 
\bea
t=T+f(\Phi), \ \ \ \phi= \Omega T +g(\Phi). 
\eea
for arbitrary functions $f$ and $g$. So we see that going to a co-rotating frame with respect to the horizon makes the time coordinate positive definite everywhere outside the horizon. The coordinate transformation  (\ref{new-transformation}) that leads to (\ref{newBTZ}) is one such choice.

That the rotating BTZ black hole is the dominant contribution to the gravity partition function was shown in \cite{Maldacena:1998bw,  Hunter}.

\section{\bf The Geodesic-Correlator Connection}

In this section, we will show that by doing a WKB approximation due to Festuccia and Liu \cite{FL1, new}, we can relate the thermal correlator postulated in the last section to spacelike geodesics that we discussed earlier. In a following section, we will identify the specific correlators that probe the Cauchy horizon. Our results are closely related to the non-rotating BTZ case, so we will proceed rather quickly. 

\subsection{Radial Scalar Field Equation}

In the case of the BTZ black hole, as mentioned in the introduction, the
major simplification is that we can solve the scalar field equation in the
metric (\ref{newBTZ}) exactly.
We first separate variables as
\eqn{separable}{\varphi=\sum_{n,\omega}\exp(ip\phi)\exp(-i\omega
t)X_{\omega,p}(r).}
The equation of
motion $(\Box-m^2)\varphi=0$ 
reduces
upon defining the lapse function
\eqn{def-N}{N^2=\frac{(r^2-r_+^2)(r^2-r_-^2)}{r^2},}
to
\eqn{X-eqnnew}{X_{\omega, p}''+\frac{(rN^2)'}{rN^2}X_{n\omega}'+\frac{1}{N^2}
\Big[\frac{\omega^2}{r^2-r_+^2}-\frac{p^2}{r^2-r_-^2}-m^2\Big]X_{\omega, p}=0.
}
We can bring this to a Schrodinger form (which will be useful when we make the WKB approximation) by defining
\eqn{schrodingerize}{Y_{\omega, p}(r)=(r^2-r_-^2)^{1/4}X_{\omega, p}(r),}
and the result can be written as
\bea
\frac{d^2 Y_{\omega, p}}{dz^2}+\Big[\omega^2-m^2(r^2-r_+^2)-\frac{p^2(r^2-r_+^2)}{r^2-r_-^2}-\frac{(3r^2+r_+^2-4r_-^2)(r^2-r_+^2)}{r^2-r_-^2}\Big]Y_{\omega, p}=0 \label{schro-form}
\eea
where we have introduced a tortoise coordinate $z$ which is defined in (\ref{newtortoise}).

\subsection{Exact Solution}

Equation (\ref{X-eqnnew}) is solvable, because the BTZ wave equation is solvable \cite{Ichinose, Keski-Vakkuri}. The wave equation in the standard BTZ coordinates is solved by using a variable redefinition of the form eqn.(57) in \cite{Keski-Vakkuri}.  We can use a similar trick here, by defining 
\eqn{ansatz}{X_{\omega,p}=(r^2-r_+^2)^{\alpha}(r^2-r_-^2)^{\beta}
Z_{\omega,p}\left(u\equiv\frac{r^2-r_-^2}{r_+^2-r_-^2}\right),}
where $Z_{\omega,p}$ is a new function with a new radial variable $u$ as
its argument, and
\begin{eqnarray}
\alpha&=&i\frac{\omega}{2\sqrt{(r_+^2-r_-^2)}}\nonumber \\
\beta&=&i\frac{p}{2\sqrt{(r_+^2-r_-^2)}}.
\end{eqnarray}
Here $u$ lies in the range $1\le u\le \infty$ as $r$ ranges between the outer horizon and the boundary.
The equation of motion now changes into the hypergeometric form
\eqn{hyper}{u(1-u)Z_{\omega,p}''+\{c-(a+b+1)u\}Z_{\omega,p}'-abZ_{\omega,p}=0,}
where
\begin{eqnarray}
a&=&\alpha+\beta+\frac{\Delta_{+}}{2} \nonumber \\
b&=&\alpha+\beta+\frac{\Delta_{-}}{2} \\
c&=&2\beta+1, \nonumber
\end{eqnarray}
with $\Delta_{\pm}$ defined in (\ref{dimensions}).

What we are interested in are solutions that satisfy finiteness-of-energy conditions appropriate for AdS.  the nature of solutions depends on whether $\nu$ in (\ref{dimensions}) is integral or not. Since we will be interested in making a large mass-expansion to connect with the geodesic approximation,
we will be interested in the large $\nu$-limit, so we can assume
that $\nu$ is not an integer. 

The general solutions that is well-defined at the boundary $(u \rightarrow \infty)$, 
\begin{eqnarray}
Z_{\omega,p}=C_1 u^{-(\alpha+\beta+h_+)}F\left(\alpha+\beta+h_+,\alpha-
\beta+h_+;1+\nu;1/u\right)+\hspace{1cm} \nonumber \\
\hspace{1cm}
+C_2u^{-(\alpha+\beta+h_-)}F\left(\alpha-\beta+h_-,
\alpha+\beta+h_-;1-\nu;1/u\right),
\end{eqnarray}
where $F \equiv {}_2 F_1$ is the hypergeometric function, and $h_{\pm}=\Delta_{\pm}/2$. The radial part
$X_{\omega,p}$ is related to the above expression
through
(\ref{ansatz}),
\eqn{aha}{X_{\omega,p}\sim (u-1)^{\alpha}u^{\beta}Z_{\omega,p}.}
Lets collect the leading behavior of the various coordinates at the boundary
and the outer horizon here for convenience:
\begin{eqnarray}
&{\rm horizon}\ (r \rightarrow r_+):& \ \ u \sim
1+\frac{r^2-r_+^2}{r_+^2-r_-^2}\rightarrow 1, \nonumber \\
&&\ z \sim
-\frac{1}{2\sqrt{(r_+^2-r_-^2)}} \ln (r^2-r_+^2) \rightarrow \infty, \hspace{1cm}
\\
&{\rm boundary}\ (r\rightarrow \infty):& \ \ u \sim
\frac{r^2}{r_+^2-r_-^2}\rightarrow \infty, \ \ z \sim
\frac{1}{r}\rightarrow 0.
\end{eqnarray}
With these, we can impose that the solution satisfy the usual AdS-falloff,
\eqn{AdS-falloff}{\phi(r,x) \rightarrow r^{-\Delta_-}\phi_0(x),}
(for some $\phi_0$) which forces  $C_2$ to be zero. Using identity (C.17) from \cite{KM} we can write
\begin{eqnarray}
X_{\omega,p}&=&C_1(u-1)^\alpha
u^{-\alpha-h_+}F(\alpha+\beta+h_+,\alpha-\beta+h_+;1+\nu;1/u) \nonumber
\\
&=&C_1\big[A(u-1)^\alpha u^\beta
F(h_++\alpha+\beta,h_-+\alpha+\beta;1+2\alpha;1-u)+ \nonumber
\\
&&\hspace{0.5cm}+B(u-1)^{-\alpha}u^{-\beta}
F(h_--\alpha-\beta,h_--\alpha-\beta;1-2\alpha;1-u)\big]
\end{eqnarray}
where
\begin{eqnarray}
A=\frac{\Gamma(1+\nu)\Gamma(-2\alpha)}{\Gamma(h_+-\alpha-\beta)
\Gamma(h_+-\alpha+\beta)},
\\
B=\frac{\Gamma(1+\nu)\Gamma(2\alpha)}{\Gamma(\alpha+
\beta+h_+)\Gamma(\alpha-\beta+h_+)}.
\end{eqnarray}
It turns out that $A=B^*$.
Using this we can see that at the horizon ($u \rightarrow 1$), the
mode $X_{\omega,p}$
reduces
to
\begin{eqnarray}
X_{\omega, p}\rightarrow
C_1|Z|\Big(
\exp{(-i\delta)}\exp(-i\omega
z)+\exp{(i\delta)}\exp(i\omega
z)\Big),\label{near-horizon}
\end{eqnarray}
where in the first line we have used the leading behavior of the
coordinates
to rewrite the expression in terms of $z$. The phase can be completely fixed, but we will not need it. 
Normalizing $X_{\omega, p}$ at the horizon with
$C_1|Z|=1$ and  using this back in the asymptotic form at the boundary,
we find that
\eqn{boundary-asymptotics}{X_{n\omega}\rightarrow
C(\omega,p)(r_+^2-r_-^2)^{(1+\nu)/2} \ r^{-(1+\nu)},}
where we have renamed $C_1$ as $C(\omega,p)$. Using the definitions of $A$ and $B$, we can explicitly write down $C(\omega,p)$:
\begin{eqnarray}
C(\omega,p)^2=
\frac{\Gamma(h_+-\alpha-\beta)\Gamma(h_+-\alpha+\beta)\Gamma(h_++\alpha+
\beta)\Gamma(h_++\alpha-\beta)}{\Gamma(1+\nu)^2\Gamma(2\alpha)\Gamma(-2\alpha)}.
\end{eqnarray}
Using ({\ref{boundarylimit}) we can now compute 
\bea
G_+(\omega,p)=\frac{4 \nu^2}{2\omega}\frac{e^{\beta \omega}}{e^{\beta \omega}-1}(r_+^2-r_-^2)^{1+\nu}C(\omega,p)^2 \hspace{1.2in} \label{G+}\\ 
=\frac{\beta (r_+^2-r_-^2)^{1+\nu} e^{\beta \omega/2}}{2\pi^2 \Gamma(\nu)^2} \Gamma(h_+-\alpha-\beta)\Gamma(h_+-\alpha+\beta)\Gamma(h_++\alpha+
\beta)\Gamma(h_++\alpha-\beta). \nonumber 
\eea
These results reduce to equations (13) and (15) in \cite{Yang:2006pc} when the black hole is static (except for some minor errors in \cite{Yang:2006pc}).

\subsection{A WKB approximation}

We can make a connection with the geodesic approximation by setting
\eqn{scaled}{p=k\nu, \ \ \omega=u\nu, \ \ m^2=\nu^2-1.}
The basic idea is that when the mass of the scalar is large (or
equivalently, $\nu \rightarrow \infty$), the
particle approximation becomes good, and so the boundary-boundary
correlator is dominated by
contributions from the bulk geodesics that connect the two points. Since the large-mass limit is a semi-classical limit, and the
scalar field equation (in the tortoise coordinate) is just a
Schr\"odinger equation with a complicated potential, the first thing we
do in order to make a connection with the geodesics is to try a WKB
solution for the ``wave-function".  Plugging in $Y=e^{\nu S}$, we get to leading order in $1/\nu$
\bea
-(\partial_z S)^2+V(z)=u^2.
\eea
where the potential
\eqn{scaled-potential}{V(z)=(r^2-r_+^2)\Big(1+\frac{k^2}{r^2-r_-^2}\Big)}
will be a crucial object in the rest of this paper. For real $k$, which will be enough for our purposes, the
plot takes the form given in figure \ref{V(z)}.
\begin{figure}
\begin{center}
\includegraphics[width=8cm]{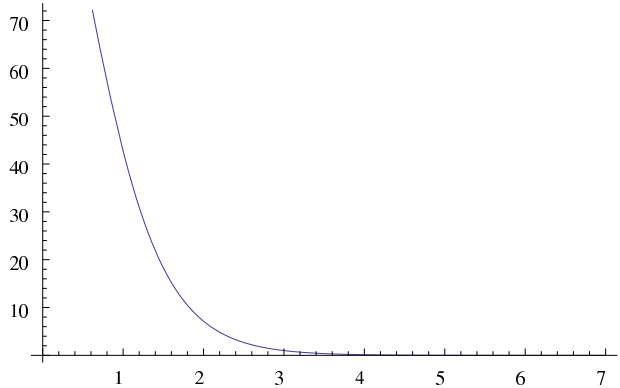}
\caption{The plot of V(z) vs. z. The outer horizon is at $z=0$ and AdS boundary at $z=\infty$.}
\label{V(z)}
\end{center}
\end{figure}
Taking advantage of the boundary condition that when the potential is infinite, the wave function is zero, we start in the classically forbidden region ($0<z<z_c$, which is the part close to the AdS boundary). Interpreting $u^2$ as the energy of the particle in the well, the standard WKB solution takes the form
\eqn{WKB1}{Y_{\omega,p}^{WKB}(z)= \frac{1}{\sqrt{\kappa(z)}}\exp(\nu
{\mathcal Z}(z))+...,}
where 
\bea
\kappa(z)=\sqrt{V-u^2}, \ \ \ {\mathcal Z}=\int_{z_c}^{z} dz' \kappa(z') 
\eea
where the turning point $z_c$ is fixed by 
\eqn{turningpoint}{(r_c^2-r_+^2)\Big(1+\frac{k^2}{r_c^2-r_-^2}\Big)=u^2.}
The dots represent higher corrections to the WKB approximation (higher in $1/\nu$). 
We start with the situation $u^2, k^2 >0$ (in which case there is a unique positive root for $r_c$) and then analytically continue to probe the regions inside the horizon.

In the classically allowed region, the standard WKB solution takes the form
\bea
Y^{WKB}_{\omega,p}(z)&=&\frac{1}{\sqrt{p(z)}}e^{i(\nu W-\frac{\pi}{4})}+cc.+..., \ \ {\rm where}\\
W (z)&=&\int_{z_c}^{z}dz' p(z'), \ \ p(z)=\sqrt{u^2-V}
\eea
Near the outer horizon, this reduces to
\bea
Y ^{WKB}_{\omega,p}(z)=\frac{1}{\sqrt{u}}e^{i(\omega z + phase)}+cc.
\eea
Comparing to (\ref{near-horizon}, \ref{schrodingerize}), we fix the relative normalizations as
\bea
Y^{WKB}_{\omega,p}(z)=\frac{Y(z)}{\sqrt{u}(r_+^2-r_-^2)^{1/4}}
\eea
Using these we find that the  boundary Wightman correlator (\ref{boundarylimit}) in the WKB limit takes the form
\bea
G_+(\omega,p)=\lim_{r,r' \rightarrow \infty}\frac{(2\nu)^2}{2\nu u}(r r')^{1+\nu} \times u \ (r_+^2-r_-^2)^{1/2} (rr')^{-1/2}Y^{WKB}_{\omega,p}(r)Y^{WKB}_{\omega,p}(r') \nonumber \\
=\lim_{r\rightarrow \infty}2\nu\sqrt{r_+^2-r_-^2}\frac{(e^{\nu {\mathcal Z}})^2}{(\sqrt{r})^2}r^{2 \nu +1} \ = \ 2\nu\sqrt{r_+^2-r_-^2} e^{\nu  Z(u,k)},\hspace{1in}
\eea
where $Z$ can be defined in terms of $z$ by
\eqn{Z}{Z=2 \lim_{z\rightarrow 0}\Big(\int_{z_c}^{z} dz' \kappa(z')-\log z\Big)}

This $Z$ can be directly related to the geodesic integrals of section 2. To see this note that ${\mathcal Z}$ satisfies
\eqn{WKB2}{\Big(\frac{d{\mathcal
Z}}{dz}\Big)^2+(r^2-r_+^2)\Big(\frac{-k^2}{r^2-r_-^2}-1\Big)=-u^2.}
This is precisely the geodesic equation (\ref{geodesic2}), if we make the
identification
\eqn{identification}{\frac{{\rm d}{\mathcal
Z}}{{\rm d}z}=-\frac{(r^2-r_-^2)^{1/2}}{r}\ \frac{{\rm d}r}{{\rm d}\lambda}, \ \
u=iE, \ \
k=iq.}
The upshot of all this is that we now have a prescription for associating
a geodesic with a given Green function determined by $Z(u,k)$: namely, for
a given $Z(u,k)$, we can associate a complex geodesic that starts and
ends at $r=+\infty$, and with the integrals of motion, $E=-iu$ and
$q=-ik$. To fully fix this
identification, we also have to specify an analytic continuation because
we are starting with real
geodesics. The analytic continuation will be discussed later: in the BTZ case it is straightforward because we also know the exact solution.

Directly integrating (\ref{WKB2}), we can write a useful expression for
$Z$:
\begin{eqnarray}
Z(u,k)&=&2\lim_{r\rightarrow\infty}({\mathcal Z}+\log
r)=2\lim_{r\rightarrow\infty}\Big(\int_{z(r_c)}^{z(r)}\frac{{\rm
d}{\mathcal
Z}}{{\rm d}z}{\rm d}z+\log r\Big) \nonumber \\
&=&2\lim_{r\rightarrow\infty}\Big(
\int_{r_c}^{r}\frac{r{\rm
d}r}{(r^2-r_+^2)\sqrt{r^2-r_-^2}}\sqrt{ (r^2-r_+^2)\Big(1+\frac{k^2}{r^2-r_-^2}\Big)-u^2}
+\log
r\Big) \nonumber \\
&=&-L(E,q)-Et(E,q)+q\phi(E,q)
\label{trans},
\end{eqnarray}
where in the last line, we have used the identification
(\ref{identification}) and the equations from section 2. This relation is a consequence of the correlator-geodesic connection. In passing, we mention that
we can interpret $L$ and $Z$ as Legendre transforms
of each other, with
\eqn{Legendre}{\frac{\partial Z}{\partial E}=-t(E,q), \ \frac{\partial
Z}{\partial q}=\phi(E,q).}
It is possible to relate the geodesic approximation to a saddle point evaluation of the coordinate space correlators, but  we will not do so here.

\section{\bf Cauchy Horizon in the Gauge Theory}

The way we probe beyind the horizon is by analytic continuation of the correlators.  Once we fix the analytic continuation, we will be able to identify the correlators that probe the Cauchy horizon using the geodesic-correlator connection.

\subsection{Analytic Continuation}

To complete the definition of $Z(u,k)$ of last section, we need to specify an analytic
continuation, both for the turning point $r_c(u,k)$ and the integration
contour. The turning point is fixed by the condition
\eqn{rc}{V(r_c)\equiv (r_c^2-r_+^2)\Big(1+\frac{k^2}{r_c^2-r_-^2}\Big)=u^2}
The positive root is
\begin{eqnarray}
r_c^2(u,k)&=&\frac{1}{2}(u^2-k^2 + r_+^2 + r_-^2 \pm
\sqrt{k^4 + 2 k^2 (r_+^2 - r_-^2 - u^2) + (r_+^2 - r_-^2 + u^2)^2}), \nonumber \\
&\equiv &
\frac{1}{2}(u^2-k^2 + r_+^2 + r_-^2\pm \sqrt{\Delta}),
\label{rcroot}
\end{eqnarray}
where we have defined a discriminant in the last line. To fully prescribe the analytic continuation, we need to fix the branch cuts. This can be done with the explicit form form of the Green function $G_+(\omega, p)$ from (\ref{G+}). It is easy to see that there is a line of poles coming from the Gamma functions, which collapse to branch cuts in the large $\nu$ limit:
\eqn{poles}{\pm i u \pm i k =\sqrt{r_+^2-r_-^2} \Big(\frac{2 n}{\nu}+1\Big).}
A plot of the analyticity struture in $u$-plane is shown in the figure. 
\begin{figure}
\begin{center}
\includegraphics[width=8cm]{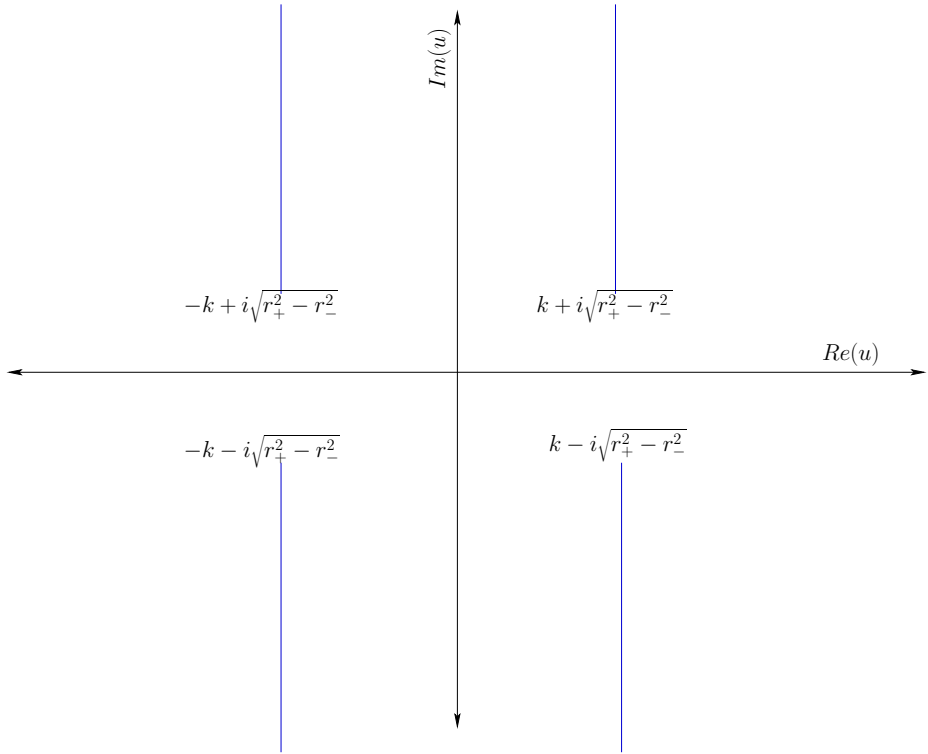}
\caption{Pole structure of $G_{+}(\omega,p)$ at large $\nu$. The branch
points are indicated.}
\label{potplot}
\end{center}
\end{figure}
With this pole structure (for $k^2>0$) its clear that we should do the analytic continuation in $u$ around the origin.  One can set the discriminant to zero and find the branch points,
\eqn{branchpoints}{u=\pm k \pm i \sqrt{r_+^2-r_-^2},}
(all combinations of signs allowed) and they agree with beginnings of the branch cuts in the figure. 

The physics of the situation is clear from the plot of the potential $U\equiv -V$ (plotted against $r$ this time) shown in figure 4. The analogy is to that of a a one-dimensional particle of energy $E^2=-u^2$ moving in a potential $-V$. The turning point gets inside the outer horizon as $u^2$ goes from positive to negative, i.e, imaginary $u$ can probe inside the outer horizon. Also, we see that for $k^2>0$, there is an infinite wall at $r=r_-$, so that no probe can get past it\footnote{Note that we have translated the problem to a {\em classical} inverse scattering problem.}. 
\begin{figure}
\begin{center}
\includegraphics[width=8cm]{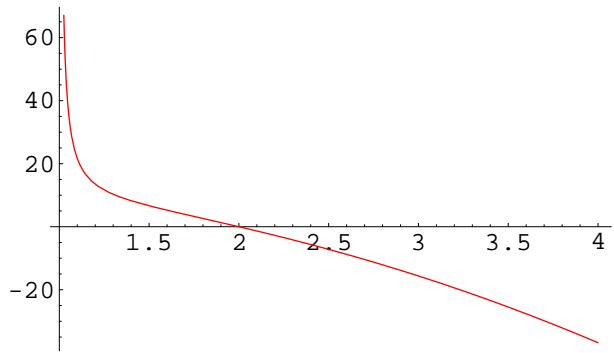}
\caption{The potential $U(r)$ (for $k^2>0$) plotted against $r$ for some choices of
parameters. The point where the
plot intersects the horizontal axis, is $r=r_+$ ($=2$ in the units
of the figure). The
inner
horizon $r=r_-$ ($=1$ in the figure)
is at the vertical axis. Notice that $u^2<0$ geodesics have positive
``energy" and can go inside the horizon.}
\label{inverse}
\end{center}
\end{figure}
Note that these statements are true for non-vanishing $k$. When $k=0$, the potential actually reduce to a flipped quadratic, and has a finite intercept on the vertical axis, meaning that for sufficiently negative $u^2$, one can reach $r_-$. But the problem is that in this case the analytic continuation that we define is not very well-defined, because the branch cuts in the u-plane (figure 3) collapse onto the imaginary axis. So we will not consider this non-generic situation in what follows. This is the caveat that was mentioned in a footnote in the introduction, and is an indication of the fact that BTZ oversimplifies the problem somewhat. A similar issue arises for static BTZ when zero-angular momentum geodesics are used as probes. In the case of the
non-rotating $AdS_5$ black hole considered by Festuccia and Liu \cite{FL1}, the
branch cuts were not parallel to the imaginary axis,
so setting $k=0$ was an acceptable simplification. In any event, in the generic case, we find that the turning point tends to the Cauchy horizon as $u \rightarrow \pm i \infty$. It should be emphasized that if one works with the usual BTZ coordinates, none of these interpretations are obvious because of a mixing of variables, and the associated lack of a clean physical picture.

Since the $u$ and $k$ variables we are working with here are rotated versions of the usual BTZ energy and (angular) momentum. The ``effective" Penrose diagram of the metric (\ref{newBTZ}) is that of a square, and in this picture, these geodesics are approximately null except for the region where they turn back at the Cauchy horizon. In effect what we are doing is to take the geodesic approximation arising from the correlator construction as the ``correct" definition of the geometry. This means that the Penrose diagram of the standard rotating BTZ black hole is effectively modified. Note that spacelike/null geodesics from the boundary can never reach the Cauchy horizon and return back in the standard Penrose diagram of the spinning BTZ black hole, figure \ref{Penrose}.

Another important point about this potential is that since it is
monotonic, it does not
allow any
(classically) bound geodesics as solutions. One way to check
this is to set
\begin{eqnarray}
\frac{dV}{dr}=0, \ \frac{d^2V}{dr^2}=0,
\end{eqnarray}
since as we tune the parameter values, at the onset of a trough in the
potential there will be a point of
inflection. The potential presented here does not admit simultaneous
solutions for these conditions. But we note that for large values of $q$ the potential does get flatter and flatter after a steep fall, even though not quite to zero-slope. These are classically quasi-stationary paths: that is, they do decay classically (and not just by tunneling), but they do so very slowly.
 
In the absence of bound geodesics BTZ is different from higher dimensional AdS black holes. There, for large
enough values of $k$, it is possible to have geodesics stuck in semi-classical equilibrium around the hole as discussed in \cite{new}. But note that the signatures of the singularity/Cauchy horizon is already visible for small $k$, so this is not a problem for our purposes.

\subsection{Signature in the Gauge Theory}

In this section, we finally put together all the ingredients and compute the correlators in the limit that they probe the Cauchy horizon. We already know from the previous subsection that $u$ going to $\pm i \infty$ is the limit where the correlators capture the Cauchy horizon. We can first compute the large-$\nu$ limit of the exact correlator (\ref{G+})\footnote{This limit can also be computed using the geodesic approximation.}, and then let $u$ go to infinity along the imaginary axis. In doing the computations of this section, we use various Sterling-like asymptotic approximations of Gamma functions along various directions in the complex plane \cite{new, Berry}.

$G_+$ reduces in the large-$\nu$ limit to
\bea
G_+(\omega,p)= 2\nu \sqrt{r_+^2-r_-^2} \ e^{\nu Z'} + ...,
\eea
where 
\bea
Z'=\frac{\beta u}{2}+A_+ \log A_++A_-\log A_-+\tilde A_+\log \tilde A_++\tilde A_-\log \tilde A_-+\log(r_+^2-r_-^2).
\eea
We have defined (in analogy with \cite{FL1}),
\bea
A_{\pm}=\frac{1}{2}\pm \frac{i(u+k)}{2\sqrt{r_+^2-r_-^2}}, \ \  \tilde A_{\pm}=\frac{1}{2}\pm \frac{i(u-k)}{2\sqrt{r_+^2-r_-^2}}
\eea
A useful fact in many of the manipulations is that $A_++A_-+\tilde A_++\tilde A_-=2$. 
The geodesic-correlator connection tells us that  $Z'$ should be equal to $Z=-E t+q \Phi-L$, once one sets $E=-i u, q=-ik$ in the latter. $Z$ can be computed by evaluating the integrals of section 2 explicitly. The result (in terms of $u, k$ is)
\bea
L(u,k)&=&-\frac{1}{2}\log (A_+ A_- \tilde A_+ \tilde A_-)-\log(r_+^2-r_-^2) \\
t(u,k)&=&\frac{1}{2}\log\Big(\frac{A_+\tilde A_+}{A_- \tilde A_-}\Big)-i\frac{\beta}{2} \\
\Phi(u,k)&=&-\frac{1}{2}\log\Big(\frac{A_+\tilde A_-}{A_- \tilde A_+}\Big)
\eea  
which correctly reproduces the expression for $Z'$.

When we let $u \rightarrow i \infty$, from the expressions above, we see that the Green function can be written as 
\bea
G_+ = \frac{2\beta}{\Gamma(\nu)^2}\Big(\frac{|\omega |}{2\sqrt{r_+^2-r_-^2}}\Big)^{2 \nu} \exp \Big(\frac{\beta(\omega - p)}{2}\Big) + .... \label{signature}
\eea
Note that in this limit $t = -i \beta/2$ and $\Phi = -i\beta/2$.  The result is quite similar to the forms written down for AdS-Schwarzschild black holes in \cite{FL1}, except that there there was an exponential fall-off associated to the fact that the Penrose diagram was not a square. In the case of the static BTZ black hole, correlators probing the singularity have the structure (\ref{signature}), except of course that the variables involved are the $\omega$ and $p$ of standard BTZ. 

\section{\bf Discussion}

In this paper, we have tried to take the viewpoint that the geodesics define the spacetime. In the large-$\nu$ limit this leads us to a picture of the Cauchy horizon for the rotating BTZ black hole that is isomorphic to that of the singularity of the static BTZ.

First we discuss the relationship of our work with earlier related work.  Levi and Ross \cite{Levi} showed that the correlators obtained by integrating over the region outside the outer horizon could be translated into correlators that involved integration over the region between the inner and outer horizons (while excluding the region inside the inner horizon). This can be taken as evidence that the CFT does not see the region behind the inner horizon. We saw how this phenomenon translates in our geodesic-based approach to the problem. Our aim is somewhat more concrete, in that we are looking for signatures of the inner horizon in the boundary correlators. We can compute the correlators that get their dominant contributions from geodesics that probe the Cauchy horizon. Our strategy is also different from the work of  \cite{Bala} where a perturbation was treated in the CFT using the analytic continuation defined for the vacuum. Their analysis is to be viewed as the breakdown of the semi-classical approximation when a perturbation is introduced (presumably tied to the fact that the inner horizon is unstable), whereas our investigations are more naturally thought of as the study of a black hole that has settled down into an equilibrium state at the end of the perturbation. 

The temperature that is associated with the construction in our work is $\sim \sqrt{r_+^2-r_-^2}$, see the appendix. This means that the extremal black hole is at zero-temperature which is again reasonable. This makes sense also from the point of entangled CFTs because Penrose diagrams for extremal black holes have only one boundary. 

The BTZ black hole is substantially simpler than higher dimensional black holes in both technical and physical aspects. At the technical level, there is the fact that wave equations in higher dimensional black holes are typically not solvable. The analytic continuation approach of Festuccia and Liu can still be applied to these black holes if we can figure out the pole structure of the Green's functions.  This can be done by computing the quasi-normal modes for Kerr-AdS black holes, an approach based on the technology presented in the appendix of \cite{FL0} might be useful for this.

Physically, the local curvature is constant for BTZ, and associated to it is the fact that the singularity is not a curvature singularity. The signature of the Cauchy horizon that we found in the correlators was related to the singularity of a non-spinning BTZ black hole, which at least for non-zero $k$ is closely related to that of higher dimensional black holes. One significant difference between static BTZ and static Schwarzschild is that in higher dimensions there is an exponential fall-off controlled by the deformation away from square shape of the Penrose diagram of the black hole\footnote{This ``non-squareness" of the Penrose diagram was discussed in \cite{Shenker2}.}. So it is tempting to speculate that even in higher dimensions, apart from such a fall-off controlled by the geometry, the signature of the Cauchy horizon in the boundary correlators will be the same as that of a singularity. But of course, a definitive claim is too premature with our analysis. 

In the case of the rotating BTZ black hole, the analyticity structure of the Green function took a simpler form in a frame where the $u$ and $k$ (or analogously $\omega$ and $p$) are rotated with respect to the standard BTZ variables. One could speculate that identifying a set of analogous variables (if they exist) might be useful in understanding the higher dimensional cases as well. The real simplification in the case of BTZ happened of course because it is a quotient, so it is an interesting question whether a co-rotating frame of some form is enough to understand higher dimensional black holes. Defining correlators with respect to a co-rotating frame seems necessary to define a QFT vacuum state in a curved background that does not run into trouble with super-radiance. It seems from this perspective that a canonical thermal state is more natural in AdS/CFT. In the canonical ensemble, one expects to see static black holes, so it is perhaps natural that the Cauchy horizon looks like a singularity. Note that from the way we construct the correlators, this statement is non-trivial - because we need the wave equation in the black hole geometry in order to define the CFT correlators, and in the black hole geometry co-rotation merely introduces a rather trivial coordinate transformation\footnote{Albeit one that changes the asymptotics, but that should hardly affect the singularity or the Cauchy horizon.}. 

Apart from the consideration of higher dimensional generalizations, one could also study charged black holes, which also have Cauchy horizons (see \cite{Brecher:2004gn}). The charged BTZ black hole is a simple possibility, it is not a quotient of AdS anymore. 

An interesting, but obviously much more difficult problem would be to study the internal structure of black rings and other exact black hole solutions discovered in higher dimensions recently \cite{5d}. Another line of investigation is to see how such constructions might be generalized to the case of recently proposed Kerr-CFT idea \cite{KerrCFT}. The problem here is that unlike in usual AdS-CFT there is no codimension one boundary in Kerr-CFT that one can usefully identify for formulating the CFT. Kerr-CFT as it is currently understood is for extremal or near-extremal black holes.

\section{\bf Acknowledgments}

I want to express my gratitude to Vijay Balasubramanian, Cyril Closset, Jarah Evslin, Guido Festuccia, Daniel Arean Fraga, Josef Lindman H\"ornlund, Hong Liu, Carlo Maccaferri, Amitabh Virmani and Chen Yang for discussions/correspondence. I am especially indebted to Hong Liu for providing me an unpublished draft of his work with Guido Festuccia. I thank my colleagues at SISSA and Solvay for friendly hospitality during much of this work.

\section{Appendix A: Causal Structure of Spacetimes}

In this appendix we review some general ideas \cite{TownsendBH} regarding the causal structure of spacetimes. The basic message is that Cauchy horizons are boundaries in spacetime beyond which predictability breaks down in general relativity. The inner horizons of rotating and charged black holes are the standard examples of Cauchy horizons. For concreteness we will deal with future Cauchy horizons, analogous definitions apply for the past as well.

We will start by defining a Cauchy surface. A {\bf (partial) Cauchy surface}, $\Sigma$, is a subset of spacetime $M$ which intersects no causal curve (i.e., spacelike or timelike curve) more than once. Roughly, this represents a spatial slice which is an instant of time. But note that this need not be a ``global" spatial slice of the spacetime. Now, a causal curve is said to be {\bf past-inextendible}, if it has no past endpoints in $M$. See figure \ref{curves} for a depiction of these essentially trivial definitions.
\begin{figure}
\begin{center}
\includegraphics[
height=0.3\textheight
]{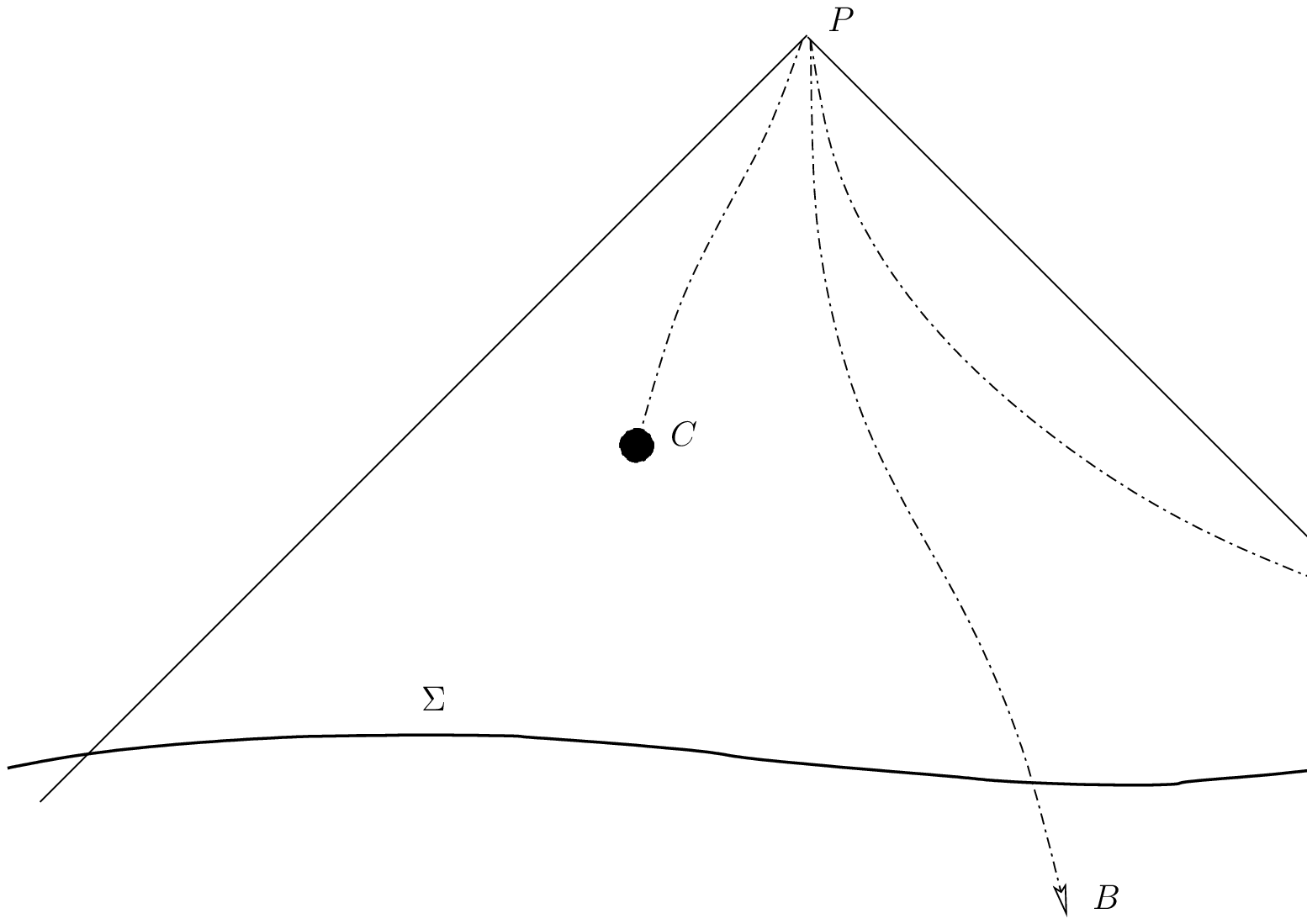}
\caption{A (partial) Cauchy surface $\Sigma$, and various kinds of curves. The past light cone of a point in the future of $\Sigma$ is shown to clarify the nature of these curves. $OA$ is not a causal curve because it gets outside the light cone, $OB$ and $OC$ are causal curves, but $OC$ is not past-inextendible because it can be continued if one chooses to.}
\label{curves}
\end{center}
\end{figure}

With these basic notions at hand, we are ready to define a Cauchy development. The {\bf future Cauchy development} of $\Sigma$, ${\cal D}^+(\Sigma)$, is comprised of the set of points $p$ in $M$ such that all past inextendible curves through $p$ intersect $\Sigma$.
\begin{figure}
\begin{center}
\includegraphics[
height=0.3\textheight
]{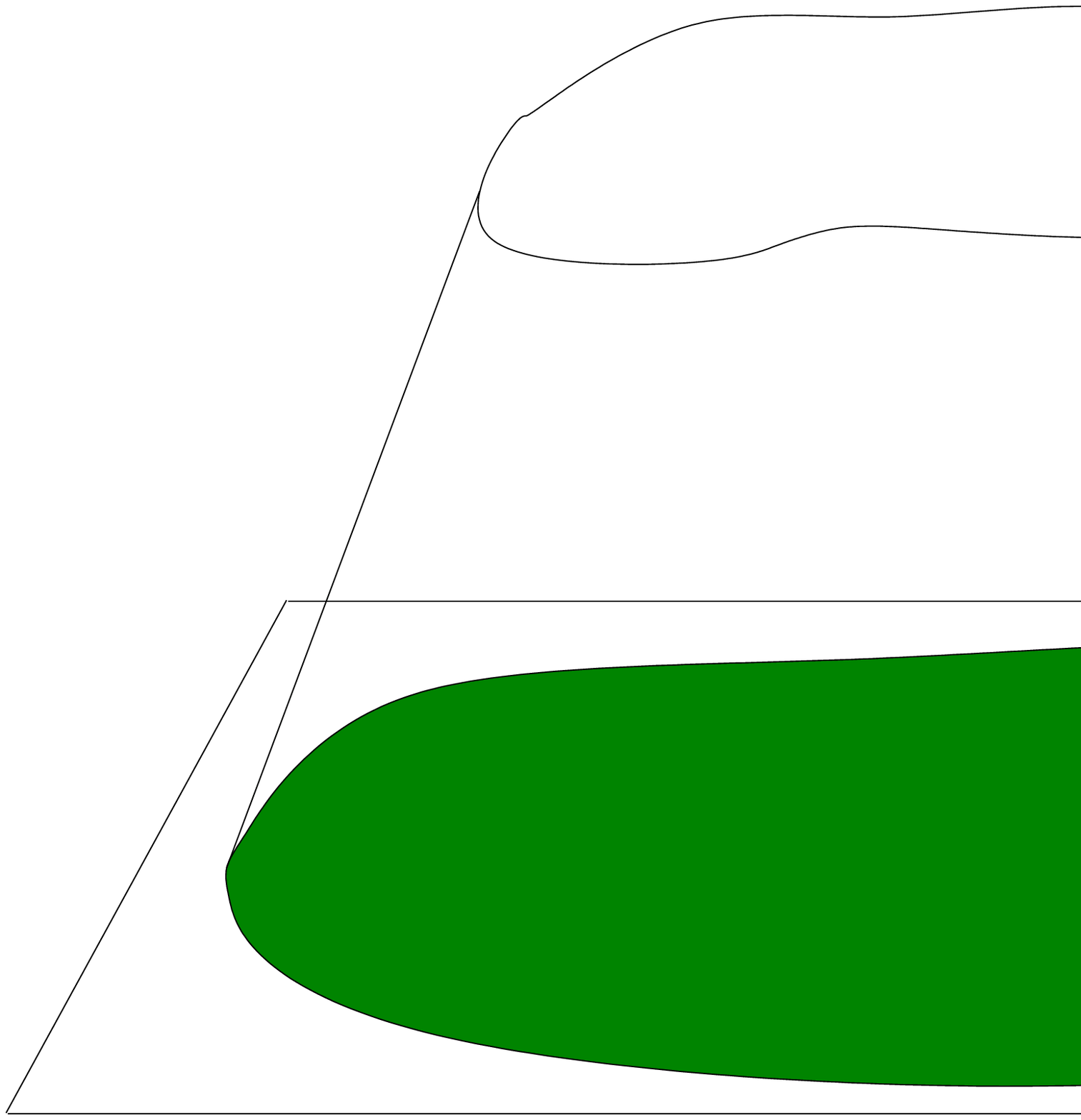}
\caption{A schematic picture of the future Cauchy development of $\Sigma$. The frustum between the parallel planes with null boundaries is a subset of ${\cal D}^+(\Sigma)$. The full ${\cal D}^+(\Sigma)$ goes on into the future. In drawing this figure we are assuming that the spacetime is roughly flat, for illustration. The structure can be more complicated in general, see eg., the BTZ Penrose diagram in the text. }
\label{curves}
\end{center}
\end{figure}
The future Cauchy development is interesting because solutions of hyperbolic PDEs (Einstein's equations) on ${\cal D}^+(\Sigma)$ are fully determined by data on $\Sigma$. We can have a similar definition for past Cauchy development. The important thing to keep in mind is that the Cauchy development is fixed {\em completely} by the data on the Cauchy surface. This should be contrasted to a future light cone. The frustum ``expands"  for the light cone as we move to the future from $\Sigma$, while it ``shrinks" for the Cauchy development.

We say that $\Sigma$ is a {\bf (global) Cauchy surface} iff ${\cal D}^+(\Sigma) \cup {\cal D}^-(\Sigma)=M$. A spacetime that admits a global Cauchy surface is called {\bf  globally hyperbolic}. This essentially means that the entire spacetime can be determined by providing Cauchy data on some surface. If $M$ is not globally hyperbolic, then $ {\cal D}^+(\Sigma)$ and/or $ {\cal D}^-(\Sigma)$ will have a boundary in $M$. This boundary is the future/past {\bf Cauchy horizon}.

Collapsing dust, maximally extended Schwarzschild and FRW cosmology are examples of globally hyperbolic spacetimes. On the other hand, the maximal extensions of charged and/or rotating black holes have Cauchy horizons, see the BTZ Penrose diagram given in Appendix B. This is because regions beyond the inner horizon are not just determined by the initial data on some Cauchy surface, but also by the boundary conditions we put on the timelike singularity. By analytic extension from the original metric, we can extend the Penrose diagram of these black holes indefinitely. We can escape to the asymptotic infinity of another universe by starting from ours, using only timelike trajectories, if we go through the Cauchy horizon. But to launch a rocket that will actually accomplish this, the initial (Cauchy) data in this Universe is not enough. The rocket will also be affected by the boundary conditions at the timelike singularity. The blueshift instability that was mentioned in the introduction is plausible from the Penrose diagram, because the entire past is visible from every point of the Cauchy horizon. Stress tensors diverge at the Cauchy horizon. Another interesting observation is that the timelike singularity is naked. This might seem like a violation of Cosmic Censorship, but it is not quite, if we require that only {\em generic} spacetimes are required to satisfy censorship. Blueshift instability makes spacetimes with Cauchy horizons non-generic against perturbations.

\section{Appendix B: BTZ Geometry}

\noindent
{\bf $AdS_3$ Quotient and BTZ Coordinates:} One of the reasons why the BTZ metric is so accessible to computations is
because it can be obtained as a quotient of $AdS_3$ by a discrete
subgroup\footnote{A
discrete isometry is an isometry which takes point P to the point
$exp(s\xi)$P, with $s=2\pi n$ for $n \in \IZ$, and $\xi$ is the Killing
vector that generates the
isometry. The claim is that with a
specific choice of $\xi$, we can get the BTZ spacetime as a
quotient of $AdS_3$ under the group $\{ exp(s\xi): s\in 2\pi \IZ \}$.
We
could write down the specific choice of $\xi$ in the standard $AdS_3$
coordinates and in principle we have the BTZ solution, but we will
first choose
a set of co-ordinates where the black hole is more intuitive and
tractable.
}
of
the isometry group $SO(2,2)$, which means that the space is locally the
same as $AdS_3$.
$AdS_3$ is defined as the hyperboloid
$-T_1^2-T_2^2+X_1^2+X_2^2=-\Lambda^2$ embedded in four-dimensional flat
space with metric $ds^2=-dT_1^2-dT_2^2+dX_1^2+dX_2^2$, where the metric on
the
hyperboloid is induced from its ambience. From here on, we set
$\Lambda=1$.
Since the black-hole is locally just AdS, we can use these embedding
coordinates to introduce a convenient coordinate system where the
quotienting operation is easy to implement.
With the advantage of hindsight \cite{BHTZ,
Levi},
we
can write,
\begin{eqnarray}
T_1&=&\sqrt{u} \cosh (r_+ \phi - r_- t) \\
T_2&=&\sqrt{u -1} \sinh(r_+ t -r_- \phi) \\
X_1&=&\sqrt{u}\sinh(r_+ \phi -r_- t) \\
X_2&=&\sqrt{u-1} \cosh (r_+ t - r_- \phi)
\end{eqnarray}
where
\eqn{alpha}{u=\frac{r^2-r_-^2}{r_+^2-r_-^2},}
and
\eqn{rplusminus}{r_{\pm}^2=\frac{M}{2}\Big[1\pm\Big(1-\frac{J^2}{M^2}
\Big)^{1/2}\Big].}
Here $M$ and $J$ will be called the mass and the angular momentum of
the black hole, with $M=r_+^2+r_-^2$ and $J=2r_+r_-$. The metric in this
system looks like,
\eqn{BTZmetric}{ds^2=-\frac{(r^2-r_+^2)(r^2-r_-^2)}{r^2}
dt^2+\frac{r^2}{(r^2-r_+^2)(r^2-r_-^2)}dr^2+r^2
\Big(d\phi-\frac{r_-r_+}{r^2}dt\Big)^2.}
Of course, the above expression
is not quite the BTZ black hole yet. It is just another coordinate system
for
(a patch of) $AdS_3$, because all we have done is introduce new
coordinates. But the advantage of this coordinate system is that here the
quotienting that we mentioned earlier is trivially implemented with an
obvious physical interpretation, as
\eqn{quotient}{\phi \sim \phi + 2 \pi.}
Therefore, with this identification understood, the above metric is the
final form
of the BTZ
black hole, and we will refer to these coordinates as the BTZ coordinates.

\noindent
{\bf Kerr-AdS Coordinates:} Even though we have not emphasized it in the text, it seems likely that the Kerr form of the BTZ metric might be useful for attempting generalizations to higher dimensions. We present an explicit cooridnate transformation and a form of the metric that is useful for comparison with higher dimensional Kerr-AdS black holes. The coordinate transformation required to do this is
\bea
t'=t, \ \phi'= \phi-a t, \ r'=\sqrt{r^2 \Sigma -a^2-\frac{2 m a^2}{\Sigma}},
\eea
with
\bea
\Sigma=1-a^2, \ r_+ r_-= \frac{2a m}{(a^2-1)^2}, \ r_+^2+r_-^2= \frac{2 a^2(1+m)-a^4-(1-2m)}{(a^2-1)^2}.
\eea
We have defined the new parameters $a$ and $m$ implicitly in terms of the original $r_{\pm}$ to avoid too much clutter. After this change of coordinates the metric takes the Boyer-Lindquist form (suppressing the primes)
\bea
ds^2=-\frac{\Delta}{r^2}\left(dt-\frac{a}{\Sigma}d\phi\right)^2+\frac{1}{r^2}\left(a dt - \frac{r^2+a^2}{\Sigma}d\phi\right)^2+\frac{r^2dr^2}{\Delta},
\eea
with
\bea
\Delta=r^4+(a^2+1-2m)r^2+a^2.
\eea

\noindent
{\bf Kruskal coordinates:} We will define Kruskal coordinates for the region around the event horizon\footnote{Throughout this paper, our philosophy is that the region outside the event horizon is the most reliable part of the geometry because among other things, the CFT can probe it with timelike geodesics. In this region, the usual ideas of AdS/CFT apply without any new subtleties: the bulk-time here is the same as the boundary time. This should be contrasted with region inside the horizon where the ``time" coordinate is spacelike and time is holographically generated.}. We write down the required coordinate transformation in the $r_+<r<\infty$ region implicitly \cite{BHTZ}:
\bea
U=\exp(-k_+ u), \ &V=\exp(k_+ v),& \Phi=\phi-\Omega_H t\ \ {\rm with} \\
&u=t-z, \ \ \ \ v= t+z&,
\eea
where $\Omega_H=r_-/r_+$ and $z=z(r)$ is the tortoise coordinate introduced in the main text. The BTZ Kruskal metric in this patch takes the form,
\bea
ds^2=-\Omega^2 dU dV +r^2\left(d\Phi+\frac{r_-}{r_+r^2}(r^2-r_+^2)\right)^2
\eea
where
\bea
\Omega^2=\frac{r_+^2(r^2-r_-)^2(r+r_+)^2}{(r_+^2-r_-^2)^2r^2}\left(\frac{r-r_-}{r+r_+}\right)^{r_-/r_+}.
\eea
{\bf New BTZ coordinates:} Two related forms of the new BTZ metric are presented in (\ref{newBTZ}) and (\ref{staticform}).

\subsection{Conformal Structure and Complexification}

We will discuss the Penrose diagram of the BTZ black hole in this section. It is important to have this in one's mind, even though in effect our geodesic construction leads us to a different (auxiliary) picture of the the geometry as seen from the CFT in the large-$\nu$ limit. 

The embedding that we defined via ``BTZ coordinates" in the previous subsection 
represents the exterior of
the black hole, as clear from the fact that $u$ and $1-u$ both
need to be
positive for it to be well-defined. But the final form of the metric is
a valid solution even in other regions. We can keep the
embedding equations to be the same in different regions, and think of the
different regions as excursions of the coordinates into the complex plane.
In other words, these different regions arise as sections of the
complexified BTZ coordinates, and in each of these sections, the metric
is real. Before writing down a prescription for going from one
region to the other, we first describe the various regions.

Notice first that the BTZ spacetime has an obvious symmetry of rotations
around $\phi$, and the
norm of the associated Killing vector $\xi=\partial_\phi$ (the
squared radius $r^2$ of the $\phi$-circle) can be used to
characterize various causal regions. When $\xi.\xi>r_+^2$, we are
outside the event horizon (region 1), when $r_-^2<\xi.\xi<r_+^2$, we are
between the Cauchy horizon and the
event horizon (region 2), and when $0<\xi.\xi<r_-^2$, we are between the
Cauchy horizon and the singularity (region 3). Region 3$'$, defined
by $-\infty<\xi.\xi<r_-^2$
contains closed time-like curves\footnote{In this region the Killing
vector $\xi$ is time-like when $\xi.\xi$ becomes negative, and so the
periodic identification immediately results in closed time-like curves.},
so we will excise the $\xi.\xi<0$ region away and consider only regions 1,
2 and 3. We can describe these different regions in
terms of the embedding coordinates as well, by noticing that
$T_1^2-X_1^2=u$ and $T_2^2-X_2^2=1-u$, and remembering the
definition of $u$:
\begin{eqnarray}
{\rm Region\ 1:} \ T_1^2-X_1^2\ge 0, \ \ T_2^2-X_2^2\le 0  \\
{\rm Region\ 2:} \ T_1^2-X_1^2\ge 0, \ \ T_2^2-X_2^2\ge 0 \\
{\rm Region\ 3':} \ T_1^2-X_1^2\le 0, \ \ T_2^2-X_2^2\ge 0
\end{eqnarray}
\begin{figure}
\begin{center}
\includegraphics[
height=0.9\textheight
]{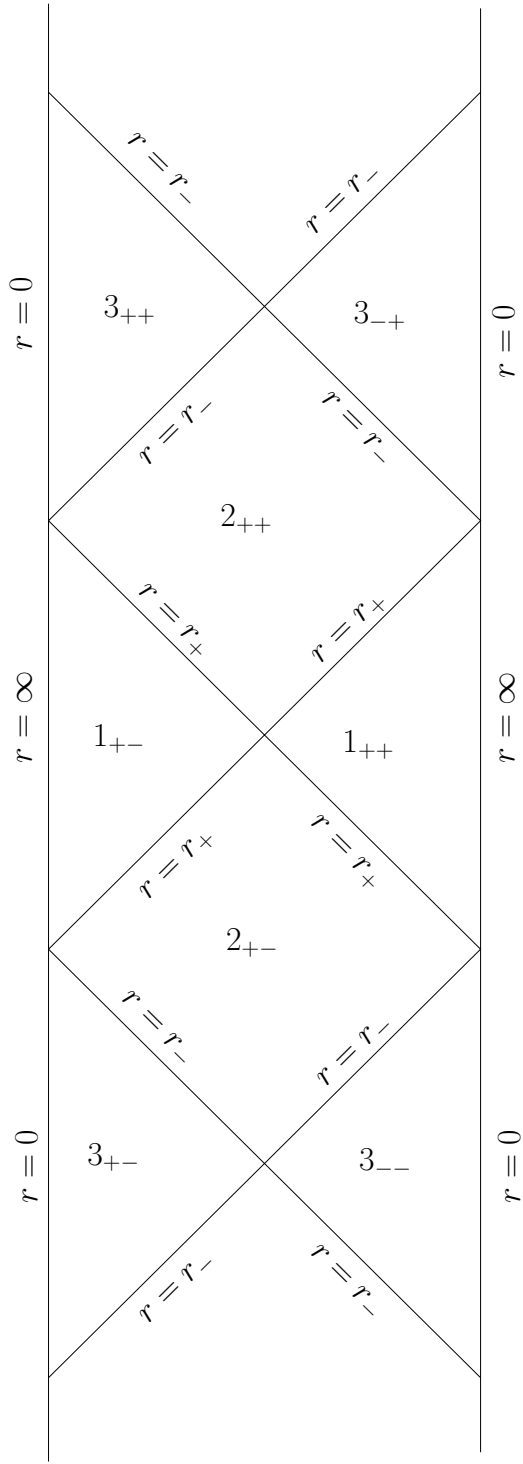}
\caption{Penrose diagram for the rotating BTZ black hole. 
The inner horizon $r=r_{-}$, the outer horizon $r=r_{+}$, the singularity
$r=0$ and the asymptotic boundary $r=\infty$ are indicated. Figure taken essentially from \cite{Bala}.}
\label{Penrose}
\end{center}
\end{figure}


As clear from the Penrose diagram, each
of the three regions in the maximally extended spacetime is composed
of several disconnected components, and these components can be labelled
by the signs of $T_1+X_1$ and $T_2+X_2$ as done in
\cite{Hemming:2002kd, Shenker1}. The different regions can be denoted
by $A_{\eta_1
\eta_2}$, where $A=1,2,3$, and $\eta_{1,2}=\pm$ are the signs of
$T_1+X_1$ and $T_2+X_2$.

The various regions in the Penrose diagram are defined above using
the original $AdS$ embedding coordinates.  To see what regions they correspond t
that end, we first note that regions 1, 2 and 3$'$ corespond
(respectively)
to $u\ge 1$, $0\le u\le 1$ and $u\le 0$. To fix the ${\pm}$-indices of
these
regions, note that
\begin{eqnarray}
T_1+X_1&=& \sqrt{u}\ \exp(r_+\phi-r_-t) \label{first1}\\
T_2+X_2&=&\sqrt{u-1}\ \exp(r_+t-r_-\phi). \label{second1}
\end{eqnarray}
We start with the physical asymptotic region $1_{++}$, where $u\ge 1$ and
$t$ is real. As we cross the horizon and move over to the region $2_{++}$,
$u$ takes the range $0\le u \le 1$ as expected since we are in region 2.
But that implies that to keep the signs of $T_1+X_1$ and $T_2+X_2$ from
(\ref{first1}), (\ref{second1}) positive, we need to shift $\phi$ and $t$
to compensate for the extra $i$:
\begin{eqnarray}
r_+\Delta t-r_-\Delta\phi=-i\frac{\pi}{2}, \\
r_+\Delta\phi-r_-\Delta t=0.\hspace{0.15in}
\end{eqnarray}
This implies that as we cross over to the new region
\begin{eqnarray}
t \rightarrow t-i\beta_+/4, \\
\phi \rightarrow \phi-i\beta_-/4,
\end{eqnarray}
where $\beta_+\equiv\beta_H=\frac{2\pi r_+}{(r_+^2-r_-^2)}$ is the inverse
Hawking temperature of the black hole, and $\beta_-=\frac{2\pi
r_-}{(r_+^2-r_-^2)}$. Similarly the cross over to another region can be
implemented as another discrete shift in the coordinates. One that will be useful to us is the crossover to the asymptotic boundary diametrically opposite to the one we started with. It is easy to see by similar arguments that to get there, we need to have
\begin{eqnarray}
t \rightarrow t-i\beta_+/2, \\
\phi \rightarrow \phi-i\beta_-/2,
\end{eqnarray}
where the original coordinates are the ones in region $1_{++}$.

If we are
interested in dealing with the transition across just the outer horizon
(or just the inner horizon for that matter), we can work in a coordinate
system whose time coordinate $t'$ is related to the BTZ time coordinate
through a $\phi$-dependent shift. See for example \cite{Bala}. The advantage of these
coordinates is that the crossover across
the outer horizon can be implemented purely by means of a $t$-shift
without doing anything to $\phi$. But
there are no coordinates where the crossovers across both the horizons can
be dealt with purely through $t$.

\section{Appendix C: Thermal Correlators}\label{state}

We define Kruskal-like coordinates by
\bea
U=\exp(-k_B u), \  \ V=\exp(k_B v), \  \   {\rm with} \ \  u=t-z, \ \  v= t+z&.
\eea
where 
\bea
k_B=\frac{1}{\sqrt{r_+^2-r_-^2}}  \ \ {\rm and} \ \ z= \frac{1}{2\sqrt{r_+^2-r_-^2}}\log\left(\frac{\sqrt{r^2-r_-^2}+\sqrt{r_+^2-r_-^2}}{\sqrt{r^2-r_-^2}+\sqrt{r_+^2-r_-^2}}\right). \label{newtortoise}
\eea
With the knowledge of $k_B$ now we can write down the Bogolubov transformation that takes us to the required thermal state \cite{Birrell, RossReview, Wald}.  The construction is standard \cite{Birrell, FL1, new}, so we present only the final result for the two-point thermal correlators here:
\bea
{\cal G}_+(r,r'; \omega, p)=\frac{1}{2\omega}\frac{e^{\beta \omega}}{e^{\beta \omega}-1}X_{\omega p}(r) X_{\omega p}(r')
\eea  
This correlator should be regarded as our definition of the appropriate thermal vacuum. The temperature can also be computed by assuming Wick rotation in $t$, and declaring regularity at the horizon. When we insert operators on opposite boundaries, the result on the right gets multiplied by $e^{-\frac{1}{2}\beta \omega}$. The $X_{\omega p}$ are the mode expansions of the scalar field (see (\ref{separable})), and
\bea
\beta=\frac{2\pi}{k_B}=\frac{2\pi}{\sqrt{r_+^2-r_-^2}},
\eea
is {\em not} the Hawking temperature of the hole. 

This gives a definition of correlators in the large-$\nu$ limit, which agrees with geodesic results.

%
\newpage

\end{document}